\documentclass[twocolumn]{aastex701}

\usepackage{amsmath}

\begin{document}

\title{Gaseous SiO and Silicate Clouds in Three Benchmark Early-L Dwarfs}

\correspondingauthor{Shu Wang}
\email{shuwang@nao.cas.cn}

\author[0009-0000-7976-7383]{Zhijun Tu}
\affiliation{CAS Key Laboratory of Optical Astronomy, National Astronomical Observatories, Chinese Academy of Sciences, Beijing 100101, People's Republic of China}
\email{zjtu@bao.ac.cn}

\author[0000-0003-4489-9794]{Shu Wang}
\affiliation{CAS Key Laboratory of Optical Astronomy, National Astronomical Observatories, Chinese Academy of Sciences, Beijing 100101, People's Republic of China}
\affiliation{School of Astronomy and Space Sciences, University of Chinese Academy of Sciences, Beijing 100049, People's Republic of China}
\email{shuwang@nao.cas.cn}

\author[0000-0001-7084-0484]{Xiaodian Chen}
\affiliation{CAS Key Laboratory of Optical Astronomy, National Astronomical Observatories, Chinese Academy of Sciences, Beijing 100101, People's Republic of China}
\affiliation{School of Astronomy and Space Sciences, University of Chinese Academy of Sciences, Beijing 100049, People's Republic of China}
\affiliation{Institute for Frontiers in Astronomy and Astrophysics, Beijing Normal University, Beijing 102206, People's Republic of China}
\email{chenxiaodian@nao.cas.cn}

\author[0009-0004-3407-0848]{Haomiao Huang}
\affiliation{CAS Key Laboratory of Optical Astronomy, National Astronomical Observatories, Chinese Academy of Sciences, Beijing 100101, People's Republic of China}
\affiliation{School of Astronomy and Space Sciences, University of Chinese Academy of Sciences, Beijing 100049, People's Republic of China}
\email{huanghm@bao.ac.cn}

\author[0000-0002-2874-2706]{Jifeng Liu}
\affiliation{CAS Key Laboratory of Optical Astronomy, National Astronomical Observatories, Chinese Academy of Sciences, Beijing 100101, People's Republic of China}
\affiliation{School of Astronomy and Space Sciences, University of Chinese Academy of Sciences, Beijing 100049, People's Republic of China}
\affiliation{Institute for Frontiers in Astronomy and Astrophysics, Beijing Normal University, Beijing 102206, People's Republic of China}
\affiliation{New Cornerstone Science Laboratory, National Astronomical Observatories, Chinese Academy of Sciences, Beijing 100012, People's Republic of China}
\email{jfliu@nao.cas.cn}

\begin{abstract}

Tracing silicon from gaseous SiO into silicate condensates provides a direct view of the condensation process that forms silicate clouds in L-dwarf atmospheres. Here we present evidence for gaseous SiO in three benchmark early-L dwarfs, HD~89744B, GJ~1048B, and LSPM~J0632+5053B, using JWST/NIRSpec and MIRI spectroscopy. Atmospheric retrievals with and without SiO under multiple cloud prescriptions identify the strongest SiO signatures at 4.0--4.2 and 7.7--8.2~$\mu$m, with the local likelihood diagnostic reaching $4.6\sigma$. In the physical cloud retrievals, the 8~$\mu$m band probes lower pressures than the 4~$\mu$m overtone, reaching atmospheric layers where equilibrium SiO is more strongly depleted by silicate condensation. All three objects also show compact Fe opacity at pressures of a few bar.  Their inferred gas-phase C/O ratios are consistent with expectations from the host-star compositions after accounting for refractory oxygen sequestration. Within the adopted Mie treatment of amorphous silicate grains, the silicate mineralogies favor Mg$_2$SiO$_4$ for HD~89744B and a mixed Mg$_2$SiO$_4$+MgSiO$_3$ cloud for GJ~1048B. Extending this framework to cooler L dwarfs will trace the coupled evolution of gaseous SiO depletion and silicate-cloud formation along the L-dwarf sequence.

\end{abstract}

\section{Introduction}\label{sec:intro}

Clouds are a defining feature of L-dwarf atmospheres. Refractory condensates form within their photospheres and strongly influence their spectra and colors. Cloud opacity produces the characteristically red near-infrared appearance of L dwarfs and modifies the atmospheric thermal structure and molecular absorption features \citep{2001ApJ...556..872A,2002ApJ...575..264T,2002ApJ...568..335M,2014A&ARv..22...80H,2015ARA&A..53..279M}. The observable properties of these clouds depend on their composition, particle size, vertical extent, and sedimentation efficiency. Variations in these properties contribute to the broad spectral and color diversity observed among L dwarfs of similar spectral type \citep{2008ApJ...686..528L,2008ApJ...678.1372C,2009ApJ...702..154S}.

\begin{deluxetable*}{lrrcccccr}[!t]
\tablecaption{Fundamental Properties of the Benchmark early-L dwarfs\label{tab:observations}}
\tabletypesize{\footnotesize}
\tablewidth{\textwidth}
\tablehead{
\colhead{Target} &
\colhead{R.A.} &
\colhead{Decl.} &
\colhead{SpT} &
\colhead{$d$} &
\colhead{Prim. SpT} &
\colhead{Age} &
\colhead{Prim. $[\mathrm{Fe/H}]$} &
\colhead{Ref.} \\
\colhead{} &
\colhead{J2000 (deg)} &
\colhead{J2000 (deg)} &
\colhead{} &
\colhead{pc} &
\colhead{} &
\colhead{Gyr} &
\colhead{dex} &
\colhead{}
}
\startdata
HD~89744B        & 155.5620 & +41.2407 & L0.5 & $38.46\pm0.74$                  & F8~IV   & 1.5--3  & $0.19\pm0.32$ & 1,2,3,7,8 \\
GJ~1048B         &  38.9997 & -23.5223 & L1.5 & $21.54\pm0.10$                  & K3.5~Vk  & 0.6--2  & $0.24\pm0.09$ & 1,4,5,7,8 \\
LSPM~J0632+5053B &  98.2021 & +50.8931 & L1.5 & $82.58\pm0.13$\tablenotemark{a} & G2 V     & 0.2--10 & $0.25\pm0.01$ & 6,7,8 \\
\enddata
\tablenotetext{a}{The adopted distance for LSPM~J0632+5053B uses the primary-star parallax.}
\tablecomments{References: (1) \citet{2014AJ....147...34S}; (2) \citet{2003AJ....126.2048G}; (3) \citet{2001AJ....122.1989W}; (4) \citet{2006AJ....132..161G}; (5) \citet{2001AJ....121.2185G}; (6) \citet{2014ApJ...792..119D}; (7) \citet{GaiaDR3_2023}; (8) \citet{2026arXiv260709851P}.}
\end{deluxetable*}

The broad 8--11~$\mu$m silicate absorption feature provides a key diagnostic of silicate-bearing cloud particles. Spitzer/IRS spectroscopy revealed broad silicate absorption in L dwarfs, attributed to small silicate grains in their atmospheres \citep{2004ApJS..154..418R,2006ApJ...648..614C}. Subsequent analyses showed that the 8--11~$\mu$m feature is strongest on average at mid-L spectral types, weakens toward the latest L dwarfs, and correlates with red near-infrared color excess at fixed spectral subtype \citep{2022MNRAS.513.5701S}. The depth, width, asymmetry, and peak wavelength of the full silicate feature are sensitive to condensate composition, particle size, and vertical cloud structure. Forward models and atmospheric retrievals have demonstrated sensitivity to several candidate silicate species, including forsterite, enstatite, and quartz, and have revealed substantial diversity in the cloud properties of young and field substellar objects \citep{2021ApJ...920..146L,2021MNRAS.506.1944B,2023MNRAS.523.4739S}. JWST observations of VHS~1256~b provide a prominent example. Its broad near- to mid-infrared spectrum shows strong silicate absorption, and recent retrievals favor vertically extended, patchy clouds containing forsterite and enstatite \citep{2023ApJ...946L...6M,Whiteford2026}. The combination of near-infrared reddening and mid-infrared silicate absorption therefore provides constraints on condensate properties over a wide wavelength range.

Gaseous SiO is expected to be an important Si-bearing species in the hot, deep layers of L-dwarf atmospheres, where temperatures remain above the silicate-condensation regime. As the atmosphere cools and Mg-bearing silicates become thermochemically stable, silicon is progressively removed from the gas phase and the SiO abundance decreases \citep{2010ApJ...716.1060V,2024ApJ...975...59M}. SiO produces a fundamental rovibrational band near 8~$\mu$m and first-overtone absorption at approximately 4.0--4.5~$\mu$m. These bands have long been observed in evolved-star atmospheres \citep{1999A&A...342..799A,2014A&A...561A..47O,2015ApJ...811...45S}. More recently, SiO has also been detected in the dayside atmosphere of the ultrahot planet WASP-121b and in the directly imaged substellar companion TWA~5~B \citep{2025NatAs...9..845E,2026MNRAS.tmp..988P}. Most relevant to L dwarfs, JWST/MIRI spectroscopy has identified 7.5--9.5~$\mu$m SiO absorption across an M5.5--L2.5 sequence, including objects with spectral types L0--L2.5 \citep{2026arXiv260715521B}. These observations make early-L atmospheres a particularly useful regime for probing the transition between gas-phase SiO and silicate condensation.

The fundamental SiO band overlaps the blue wing of the broad silicate cloud feature. Because both opacity sources shape the spectrum around 8~$\mu$m, this overlap introduces degeneracies in atmospheric retrievals. The inferred SiO abundance can therefore depend on the adopted cloud prescription, while the treatment of SiO opacity can in turn affect the retrieved cloud properties. Broad wavelength coverage helps break this degeneracy by simultaneously constraining the near-infrared cloud continuum, the 4~$\mu$m SiO overtone, and the full 8--11~$\mu$m silicate profile. Benchmark early-L dwarfs are especially valuable because their well-characterized primaries provide independent constraints on system age and bulk composition.

In this work, we investigate gaseous SiO and silicate clouds in three benchmark early-L dwarfs: HD~89744B (L0.5), GJ~1048B (L1.5), and LSPM~J0632+5053B (L1.5). Using broad JWST/NIRSpec and MIRI spectroscopy, we test whether gaseous SiO can be distinguished from overlapping silicate cloud opacity and assess how the inferred SiO signal depends on cloud treatment. We first describe the JWST observations and data reduction in Section~\ref{sec:data}, and derive bolometric luminosities and evolutionary model constraints in Section~\ref{sec:evolutionary_parameters}. We then summarize the atmospheric retrieval framework in Section~\ref{sec:methods}, and the detailed methodology and parameter priors are provided in Appendix~\ref{app:retrieval_framework}. In Section~\ref{sec:results}, we use these retrievals to examine the SiO signatures, their connection to vertical cloud structure, the atmospheric compositions relative to the host stars, and the silicate mineralogies favored by the spectra. Our conclusions are summarized in Section~\ref{sec:conclusions}.

\begin{deluxetable}{lrrrrr}
\tablecaption{\texttt{SEDkit}-derived Luminosities and Sonora Diamondback Evolutionary Predictions\label{tab:sed_diamondback}}
\tabletypesize{\small}
\tablewidth{1pt}
\tablehead{
\colhead{Target} &
\colhead{$[\mathrm{M/H}]$} &
\colhead{$\log_{10}(L_{\rm bol}/L_\odot)$} &
\colhead{$T_{\rm eff}$} &
\colhead{$\log g$} &
\colhead{$R$} \\
\colhead{} &
\colhead{dex} &
\colhead{dex} &
\colhead{K} &
\colhead{cgs} &
\colhead{$R_{\rm Jup}$}
}
\startdata
HD~89744B & $-0.5$ & $-3.65_{-0.02}^{+0.02}$ & $2283_{-13}^{+12}$ & $5.39_{-0.01}^{+0.01}$ & $0.92_{-0.01}^{+0.01}$ \\
            & $0.0$  &                              & $2226_{-17}^{+14}$ & $5.33_{-0.01}^{+0.01}$ & $0.98_{-0.01}^{+0.01}$ \\
            & $+0.5$ &                              & $2156_{-16}^{+16}$ & $5.27_{-0.01}^{+0.01}$ & $1.05_{-0.01}^{+0.01}$ \\
GJ~1048B   & $-0.5$ & $-3.69_{-0.01}^{+0.01}$ & $2241_{-16}^{+8}$ & $5.37_{-0.04}^{+0.01}$ & $0.92_{-0.01}^{+0.01}$ \\
            & $0.0$  &                              & $2175_{-16}^{+8}$ & $5.32_{-0.04}^{+0.01}$ & $0.98_{-0.01}^{+0.01}$ \\
            & $+0.5$ &                              & $2107_{-16}^{+7}$ & $5.26_{-0.04}^{+0.01}$ & $1.04_{-0.01}^{+0.02}$ \\
LSPM~J0632+5053B & $-0.5$ & $-3.69_{-0.01}^{+0.01}$ & $2245_{-86}^{+5}$ & $5.38_{-0.19}^{+0.01}$ & $0.92_{-0.01}^{+0.07}$ \\
            & $0.0$  &                              & $2178_{-86}^{+4}$ & $5.33_{-0.19}^{+0.01}$ & $0.97_{-0.01}^{+0.08}$ \\
            & $+0.5$ &                              & $2110_{-86}^{+4}$ & $5.27_{-0.19}^{+0.01}$ & $1.04_{-0.01}^{+0.09}$ \\
\enddata
\tablecomments{These quantities are predictions from the Sonora Diamondback hybrid $f_{\rm sed}=2$ evolutionary tracks. Values are medians with 16th/84th-percentile offsets.}
\end{deluxetable}

\section{JWST Observations and Data Reduction}\label{sec:data}

The sample comprises three wide, common-motion early-L companions to FGK primaries. Table~\ref{tab:observations} summarizes the properties of the companions and their host stars. All three targets were observed under JWST GO program 3670 (PI B. Burningham) with NIRSpec/PRISM and MIRI/LRS. We retrieved the raw \texttt{uncal} products from the Mikulski Archive for Space Telescopes (MAST)\footnote{\url{https://mast.stsci.edu/search/ui/\#/jwst/}} and reprocessed the NIRSpec and MIRI observations uniformly.

We processed the \texttt{uncal} files through the Detector1, Spec2, and Spec3 stages of the JWST Calibration Pipeline, version 3.0.0. Calibration reference files were managed with CRDS version 13.2.7 using the fixed context \texttt{jwst\_1581.pmap} \citep{2016A&C....16...41G}. We used the default pipeline parameters throughout. The reduction produced one NIRSpec/PRISM spectrum and one MIRI/LRS spectrum for each target. The native level-3 products cover approximately 0.55--5.38~$\mu$m for NIRSpec and 5--14.02~$\mu$m for MIRI \citep{2015PASP..127..623K,2022A&A...661A..80J}. We retained spectral points with finite fluxes and positive uncertainties.

The NIRSpec and MIRI spectra overlap between approximately 5 and 5.3~$\mu$m. A direct comparison in the overlap region shows that their flux levels agree to within 2\% for all three targets, requiring no additional multiplicative scaling between the two instruments. Because NIRSpec provides both higher spectral resolution and higher signal-to-noise ratios (S/N) over this wavelength range, we use the NIRSpec data up to 5.3~$\mu$m. The MIRI spectra show progressively lower S/N toward longer wavelengths, and we restrict the analysis to wavelengths below 12~$\mu$m. The spectra used in the subsequent analysis thus combine NIRSpec data over 0.8--5.3~$\mu$m with MIRI data over 5.3--12~$\mu$m. For LSPM~J0632+5053B, the MIRI spectrum becomes noise dominated beyond approximately 9.3~$\mu$m, and we further restrict its wavelength coverage to 0.8--9.3~$\mu$m.

\section{Bolometric Luminosities and Evolutionary Model Predictions}\label{sec:evolutionary_parameters}

We used \texttt{SEDkit} \citep{2015ApJ...810..158F} to derive the bolometric luminosity of each companion from its observed spectrum, following the procedure of \citet{2015ApJ...810..158F}. We then combined the resulting luminosity distributions with the literature age constraints listed in Table~\ref{tab:observations}, sampling the ages uniformly in logarithmic space, and propagated them through the Sonora Diamondback hybrid evolutionary tracks \citep{2024ApJ...975...59M}. At each of $[\mathrm{M/H}]=-0.5$, 0.0, and $+0.5$, we used Monte Carlo sampling to derive distributions of effective temperature, surface gravity, and radius. The resulting bolometric luminosity and evolutionary model estimates are summarized in Table~\ref{tab:sed_diamondback}.

The evolutionary predictions are used to motivate the priors on \(\log g\) and radius adopted in the retrievals. Similar evolution-informed constraints on bulk parameters have been incorporated into retrieval analyses of directly imaged and isolated substellar objects \citep{2023AJ....166..198Z,2025AJ....170...64Z,2025A&A...703A..79M}.

\section{Atmospheric Retrieval Framework}
\label{sec:methods}

We fit HD~89744B, GJ~1048B, and LSPM~J0632+5053B independently. For each target, we consider three baseline cloud treatments: cloud-free, gray, and a physical condensate cloud model containing Mg$_2$SiO$_4$ and Fe. Each cloud treatment is retrieved with and without gaseous SiO, giving six baseline models per target and 18 across the full sample. 

To examine the silicate mineralogy favored by the spectra, we also perform retrievals with alternative cloud compositions, including MgSiO$_3$, Mg$_2$SiO$_4$+MgSiO$_3$, and MgSiO$_3$+SiO$_2$. These models use the same retrieval framework as the baseline physical cloud model and are described in Appendix~\ref{sec:clouds}.

\setlength{\tabcolsep}{8pt}
\begin{deluxetable*}{lccccccc}
\tablecaption{Atmospheric Retrieval Results and Within-Source Model Ranking\label{tab:retrieval_results}}
\tabletypesize{\scriptsize}
\tablewidth{0pt}
\tablehead{
\colhead{Model} &
\colhead{$T_{\rm eff}$} &
\colhead{$\log g$} &
\colhead{$R$} &
\colhead{$[\mathrm{M/H}]_{\rm gas}$} &
\colhead{$(\mathrm{C/O})_{\rm gas}$} &
\colhead{$\log f_{\rm SiO}$} &
\colhead{$\Delta\mathrm{BIC}$} \\
\colhead{} &
\colhead{K} &
\colhead{cgs} &
\colhead{$R_{\rm Jup}$} &
\colhead{dex} &
\colhead{} &
\colhead{} &
\colhead{}
}
\startdata
\cutinhead{HD~89744B}
Mg$_2$SiO$_4$+Fe+SiO & $2168_{-6}^{+6}$ & $5.25_{-0.03}^{+0.04}$ & $1.03_{-0.02}^{+0.02}$ & $0.46_{-0.04}^{+0.03}$ & $0.76_{-0.03}^{+0.02}$ & $-0.17_{-0.08}^{+0.08}$ & $0$ \\
Mg$_2$SiO$_4$+Fe & $2164_{-8}^{+23}$ & $5.27_{-0.03}^{+0.03}$ & $1.04_{-0.03}^{+0.02}$ & $0.47_{-0.03}^{+0.03}$ & $0.73_{-0.02}^{+0.02}$ & $\cdots$ & $52$ \\
Gray & $2239_{-6}^{+6}$ & $5.18_{-0.05}^{+0.06}$ & $0.97_{-0.02}^{+0.02}$ & $0.53_{-0.02}^{+0.02}$ & $0.69_{-0.05}^{+0.04}$ & $\cdots$ & $144$ \\
Gray+SiO & $2239_{-6}^{+6}$ & $5.18_{-0.05}^{+0.06}$ & $0.97_{-0.02}^{+0.02}$ & $0.51_{-0.02}^{+0.02}$ & $0.69_{-0.05}^{+0.04}$ & $-2.42_{-0.36}^{+0.45}$ & $150$ \\
Cloud-free & $2253_{-7}^{+7}$ & $5.26_{-0.07}^{+0.08}$ & $0.96_{-0.02}^{+0.02}$ & $0.52_{-0.07}^{+0.04}$ & $0.60_{-0.03}^{+0.02}$ & $\cdots$ & $165$ \\
Cloud-free+SiO & $2252_{-7}^{+7}$ & $5.26_{-0.07}^{+0.08}$ & $0.96_{-0.02}^{+0.02}$ & $0.51_{-0.07}^{+0.04}$ & $0.60_{-0.03}^{+0.02}$ & $-2.30_{-0.45}^{+0.55}$ & $171$ \\
\cutinhead{GJ~1048B}
Mg$_2$SiO$_4$+Fe+SiO & $2127_{-5}^{+5}$ & $5.16_{-0.04}^{+0.04}$ & $1.02_{-0.01}^{+0.01}$ & $0.45_{-0.02}^{+0.01}$ & $0.89_{-0.03}^{+0.02}$ & $-0.20_{-0.10}^{+0.10}$ & $0$ \\
Mg$_2$SiO$_4$+Fe & $2132_{-7}^{+8}$ & $5.22_{-0.04}^{+0.04}$ & $1.01_{-0.01}^{+0.01}$ & $0.48_{-0.02}^{+0.01}$ & $0.85_{-0.02}^{+0.02}$ & $\cdots$ & $49$ \\
Gray+SiO & $2180_{-8}^{+7}$ & $5.23_{-0.06}^{+0.07}$ & $0.97_{-0.01}^{+0.01}$ & $0.48_{-0.03}^{+0.02}$ & $0.79_{-0.04}^{+0.04}$ & $-0.53_{-0.11}^{+0.12}$ & $104$ \\
Gray & $2176_{-6}^{+7}$ & $5.21_{-0.05}^{+0.07}$ & $0.98_{-0.01}^{+0.01}$ & $0.47_{-0.03}^{+0.02}$ & $0.85_{-0.03}^{+0.02}$ & $\cdots$ & $140$ \\
Cloud-free+SiO & $2197_{-5}^{+5}$ & $5.28_{-0.07}^{+0.07}$ & $0.96_{-0.01}^{+0.01}$ & $0.50_{-0.03}^{+0.02}$ & $0.71_{-0.02}^{+0.02}$ & $-0.50_{-0.09}^{+0.08}$ & $146$ \\
Cloud-free & $2199_{-6}^{+6}$ & $5.22_{-0.09}^{+0.11}$ & $0.96_{-0.01}^{+0.01}$ & $0.47_{-0.03}^{+0.02}$ & $0.84_{-0.07}^{+0.07}$ & $\cdots$ & $207$ \\
\cutinhead{LSPM~J0632+5053B}
Gray+SiO & $2160_{-5}^{+5}$ & $5.18_{-0.05}^{+0.05}$ & $0.99_{-0.01}^{+0.01}$ & $0.45_{-0.03}^{+0.03}$ & $0.82_{-0.02}^{+0.02}$ & $-0.11_{-0.11}^{+0.10}$ & $0$ \\
Mg$_2$SiO$_4$+Fe+SiO & $2169_{-5}^{+5}$ & $5.22_{-0.04}^{+0.04}$ & $0.98_{-0.01}^{+0.01}$ & $0.46_{-0.02}^{+0.02}$ & $0.82_{-0.02}^{+0.01}$ & $-0.16_{-0.10}^{+0.11}$ & $18$ \\
Mg$_2$SiO$_4$+Fe & $2159_{-7}^{+8}$ & $5.21_{-0.03}^{+0.03}$ & $0.99_{-0.01}^{+0.01}$ & $0.48_{-0.02}^{+0.02}$ & $0.81_{-0.02}^{+0.01}$ & $\cdots$ & $24$ \\
Gray & $2171_{-5}^{+5}$ & $5.20_{-0.05}^{+0.06}$ & $0.98_{-0.01}^{+0.01}$ & $0.47_{-0.04}^{+0.02}$ & $0.82_{-0.02}^{+0.02}$ & $\cdots$ & $40$ \\
Cloud-free+SiO & $2188_{-9}^{+9}$ & $5.28_{-0.08}^{+0.08}$ & $0.96_{-0.01}^{+0.01}$ & $0.43_{-0.07}^{+0.05}$ & $0.77_{-0.05}^{+0.04}$ & $-0.08_{-0.13}^{+0.13}$ & $96$ \\
Cloud-free & $2198_{-9}^{+10}$ & $5.24_{-0.07}^{+0.09}$ & $0.96_{-0.01}^{+0.01}$ & $0.46_{-0.06}^{+0.04}$ & $0.79_{-0.04}^{+0.03}$ & $\cdots$ & $130$ \\
\enddata
\tablecomments{
Within each source, models are ordered by increasing BIC, with
$\Delta{\rm BIC}\equiv{\rm BIC}-{\rm BIC}_{\min}$.
Reported parameters are posterior medians with 16th--84th percentile uncertainties. $[\mathrm{M/H}]{\rm gas}$ and $(\mathrm{C/O}){\rm gas}$ represent the gas-phase elemental composition implied by the retrieved molecular abundances. The \texttt{+SiO} suffix denotes models including gaseous SiO.
}
\end{deluxetable*}

We compute all model spectra with \texttt{petitRADTRANS} \citep{2019A&A...627A..67M,2024JOSS....9.5875N}. Because the resolving power of the JWST spectra varies with wavelength, higher resolution regions contain more spectral samples and would carry more weight in the likelihood simply because they contain more data points. We therefore bin regions with $R>100$ to a target resolving power of $R=100$. The full retrieval framework and parameter priors are described in Appendix~\ref{app:retrieval_framework}.

\section{Results and Discussion}
\label{sec:results}

Table~\ref{tab:retrieval_results} presents the six baseline retrievals for each source, combining cloud-free, gray cloud, and Mg$_2$SiO$_4$+Fe cloud prescriptions with and without SiO hypotheses. We rank these models using
${\rm BIC}=k\ln N-2\ln\mathcal{L}_{\max}$,
where $k$ is the number of free parameters and $N$ is the number of fitted spectral data points. The reported quantity is
$\Delta{\rm BIC}={\rm BIC}-{\rm BIC}_{\min}$,
evaluated among the six baseline retrievals for each source. The additional physical cloud retrievals used to compare silicate mineralogies form a separate model set and are discussed in Section~\ref{subsec:mineralogy}.

To quantify the local preference for SiO in the two diagnostic wavelength regions, we define a sigma-equivalent local likelihood diagnostic, $S_{\rm SiO}={\rm sign}(\Delta\ln\mathcal{L}_{\rm local})\sqrt{2|\Delta\ln\mathcal{L}_{\rm local}|}$, where $\Delta\ln\mathcal{L}_{\rm local}=\ln\mathcal{L}_{\rm with,local}-\ln\mathcal{L}_{\rm no,local}$. Positive $S_{\rm SiO}$ values favor models including SiO. The local likelihoods are evaluated over the selected wavelength ranges using the maximum likelihood solutions from fits to the full spectrum, with the corresponding additive variance terms included. Figures~\ref{fig:hd_spectral}--\ref{fig:lspm_spectral} show these comparisons around the 4.0--4.2 and 7.7--8.2~$\mu$m SiO bands for the three objects.

The 4.0--4.2~$\mu$m region contains the SiO first overtone, while 7.7--8.2~$\mu$m covers the most distinctive blue side of the fundamental band. In the physical cloud fits, SiO contributes appreciable opacity throughout 7.7--9.3~$\mu$m, although the total opacity over this wavelength range is still dominated by the silicate cloud. At $R=100$, variations in the silicate cloud opacity can reproduce much of the spectral structure beyond approximately 8.2~$\mu$m, increasing the degeneracy between the two opacity sources. Across all three objects, SiO is recovered by the statistically and spectrally viable cloudy models, while simpler cloud treatments can change the inferred strength of the local SiO signature.

\subsection{HD~89744B}
\label{subsec:hd_results}

The Mg$_2$SiO$_4$+Fe+SiO retrieval is the preferred baseline model for HD~89744B, with its no-SiO counterpart at $\Delta{\rm BIC}=52$. Gray and cloud-free models are strongly disfavored ($\Delta{\rm BIC}\geq144$), supporting a physical silicate cloud model that includes SiO.

\begin{figure*}[t]
    \centering
    \includegraphics[width=\textwidth]{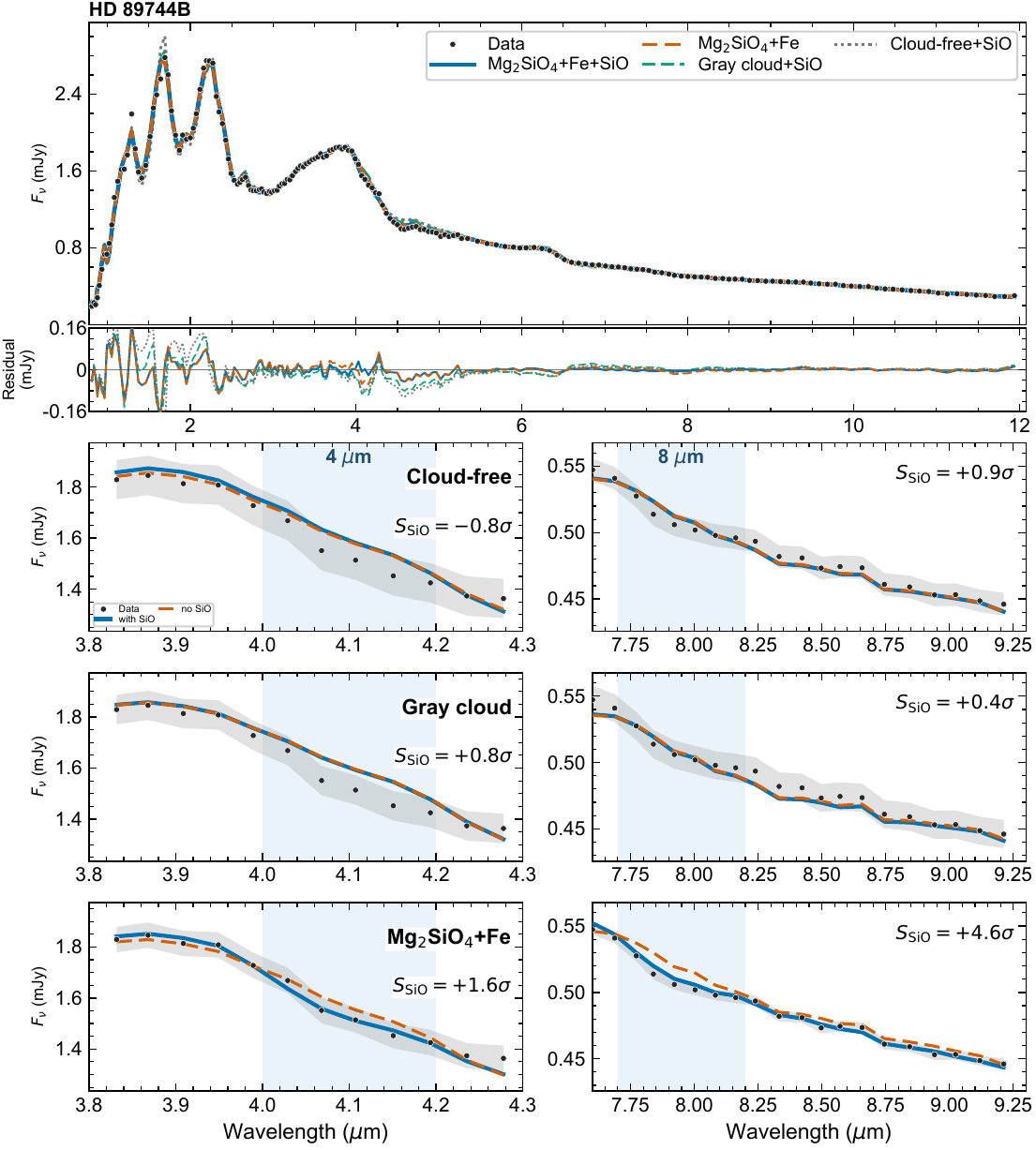}
    \caption{Spectral model comparison for HD~89744B. The upper panel shows the JWST spectrum with selected retrieval models, and the panel below shows the corresponding data-model residuals. The lower panels compare the fits with and without SiO for the cloud-free, gray cloud, and Mg$_2$SiO$_4$+Fe prescriptions around the 4 and 8~$\mu$m SiO bands. The blue shaded regions mark the 4.0--4.2 and 7.7--8.2~$\mu$m intervals used to evaluate $S_{\rm SiO}$. Gray bands show the effective $1\sigma$ flux uncertainties after inclusion of the fitted additive variance term.\label{fig:hd_spectral}}
\end{figure*}

\begin{figure*}
    \centering
    \includegraphics[width=\textwidth]{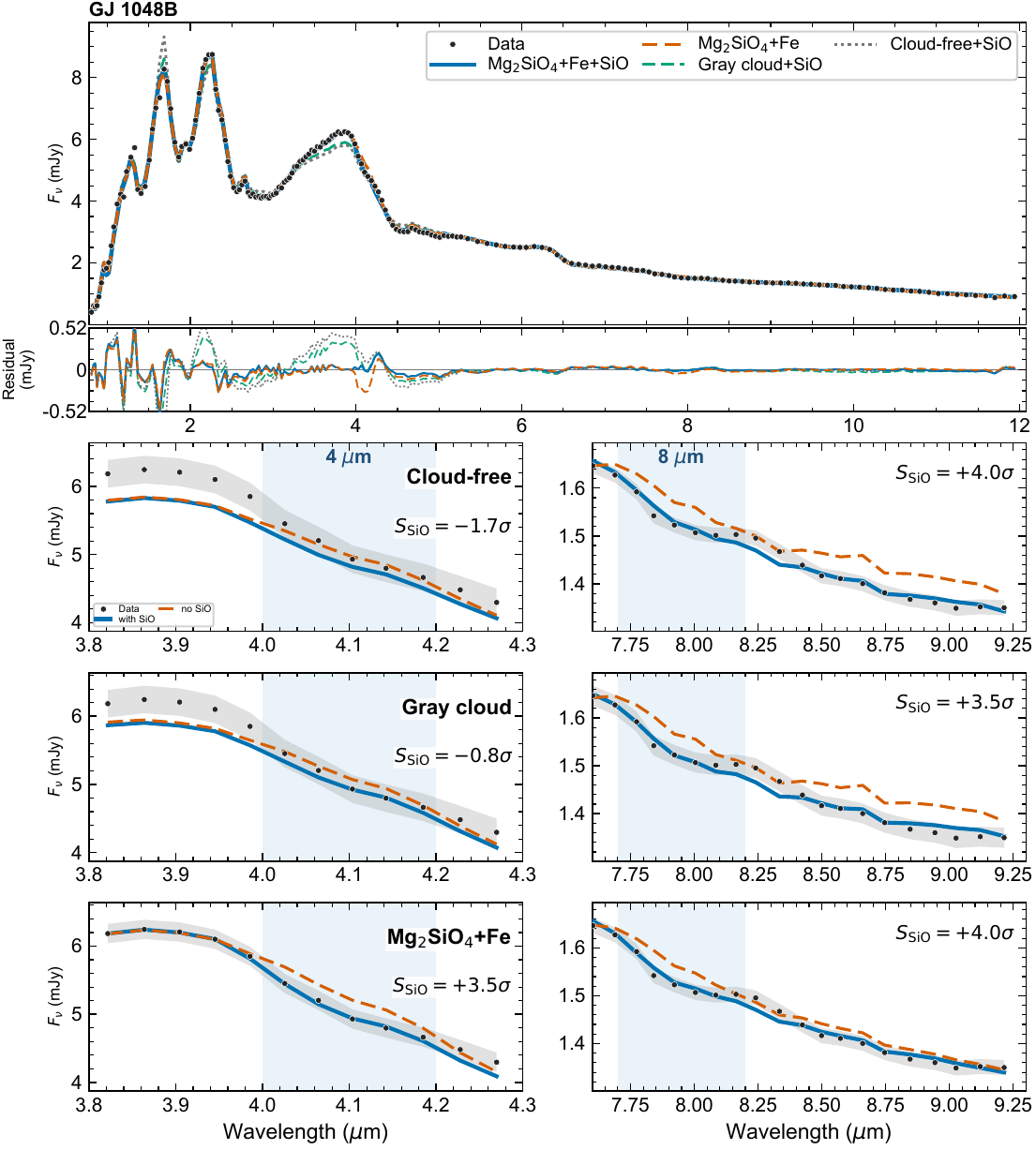}
    \caption{Same format as Figure~\ref{fig:hd_spectral}, but for GJ~1048B.\label{fig:gj_spectral}}
    \end{figure*}

Within the Mg$_2$SiO$_4$+Fe models, SiO is preferred at $+1.6\sigma$ around 4~$\mu$m and $+4.6\sigma$ around 8~$\mu$m. The stronger 8~$\mu$m preference arises from the 7.7--8.2~$\mu$m flux decline, where the no-SiO model systematically overpredicts the observed flux (Figure~\ref{fig:hd_spectral}). Beyond $\sim8.2~\mu$m, cloud-property adjustments partly compensate for the broader SiO opacity. The simplified cloud models show only weak local SiO preferences ($|S_{\rm SiO}|<1\sigma$), but are already strongly disfavored by the full-spectrum fits.

\subsection{GJ~1048B}
\label{subsec:gj_results}

\begin{figure*}
    \centering
    \includegraphics[width=\textwidth]{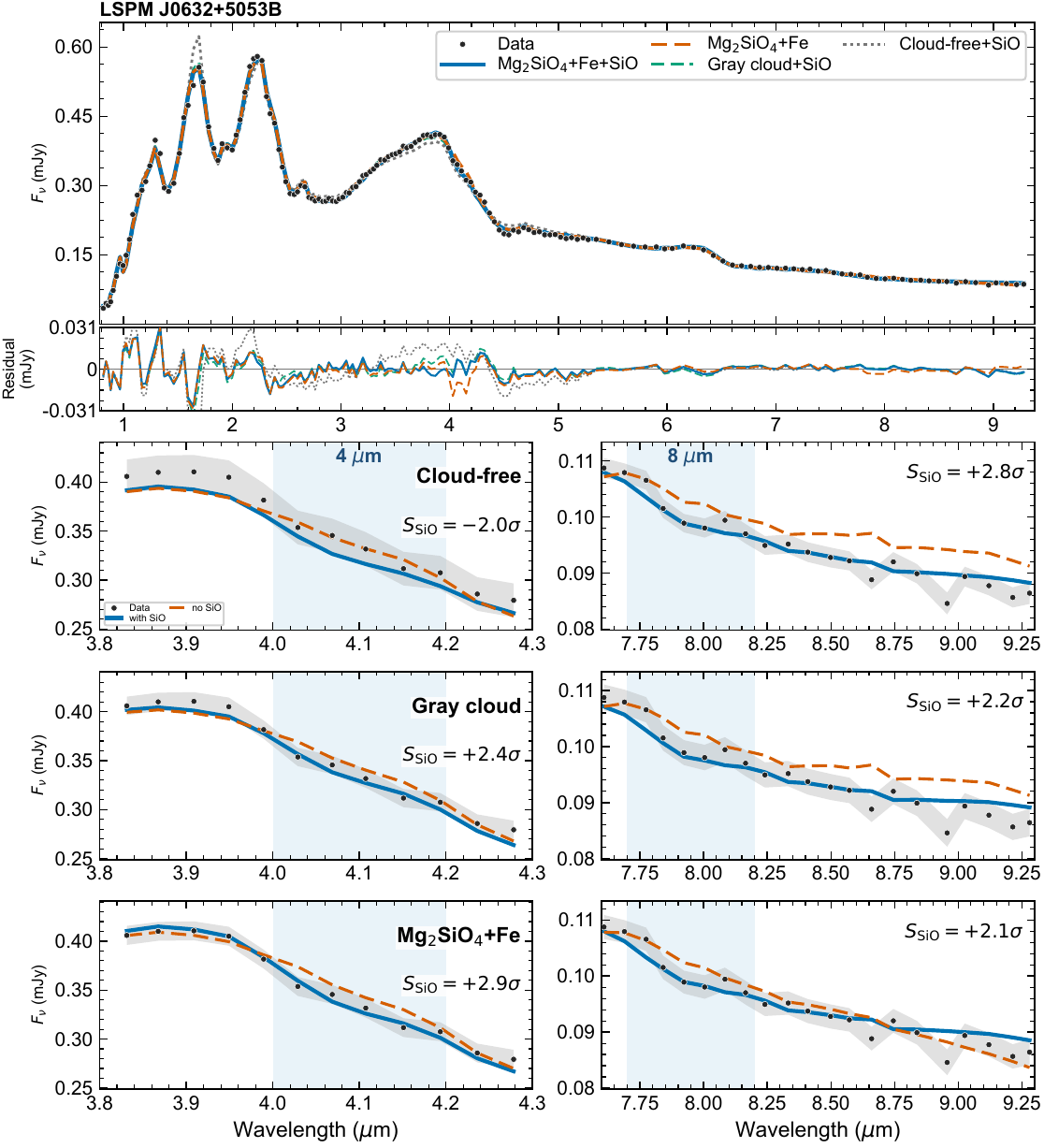}
    \caption{Same format as Figure~\ref{fig:hd_spectral}, but for LSPM~J0632+5053B.\label{fig:lspm_spectral}}
\end{figure*}

GJ~1048B shows the clearest SiO signature among the three sources. The Mg$_2$SiO$_4$+Fe+SiO model has the lowest baseline BIC, with its no-SiO counterpart at $\Delta{\rm BIC}=49$, while the gray+SiO and cloud-free+SiO models are less favored overall ($\Delta{\rm BIC}=104$ and 146, respectively).

The Mg$_2$SiO$_4$+Fe models yield consistent SiO preferences of $+3.5\sigma$ at 4~$\mu$m and $+4.0\sigma$ at 8~$\mu$m (Figure~\ref{fig:gj_spectral}). The 8~$\mu$m preference remains strong in the gray and cloud-free models ($+3.5\sigma$ and $+4.0\sigma$), whereas their 4~$\mu$m values become negative ($-0.8\sigma$ and $-1.7\sigma$). This difference likely arises from their underprediction of the 3.9~$\mu$m spectral peak, which biases the local continuum over 4.0--4.2~$\mu$m. The physical cloud model better reproduces this morphology and recovers a positive SiO preference, while the 7.7--8.2~$\mu$m feature is consistently recovered across all baseline cloud prescriptions.

\subsection{LSPM~J0632+5053B}
\label{subsec:lspm_results}

LSPM~J0632+5053B shows evidence for SiO in both cloudy model families. The gray and Mg$_2$SiO$_4$+Fe models give $S_{\rm SiO}=+2.4\sigma$ and $+2.9\sigma$ at 4~$\mu$m, and $+2.2\sigma$ and $+2.1\sigma$ at 8~$\mu$m, respectively (Figure~\ref{fig:lspm_spectral}). The cloud-free model gives $-2.0\sigma$ at 4~$\mu$m but $+2.8\sigma$ at 8~$\mu$m. Thus, similar to GJ~1048B, the 8~$\mu$m signature is recovered across all baseline cloud treatments, whereas the 4~$\mu$m result is more sensitive to the local continuum.

The gray+SiO model has the lowest baseline BIC, with the Mg$_2$SiO$_4$+Fe+SiO and no-SiO models at $\Delta{\rm BIC}=18$ and 24. This preference for the simpler cloud model likely reflects the limited wavelength coverage to 9.3~$\mu$m, which misses the red wing of the broad silicate feature and leaves the Mg$_2$SiO$_4$ component weakly constrained (Figure~\ref{fig:three_source_structure}). Consistently, truncating the HD~89744B and GJ~1048B spectra at 9.3~$\mu$m lowers the retrieved Mg$_2$SiO$_4$ abundances and broadens their posteriors.

The gray cloud can reproduce the broad opacity with fewer parameters and mimic much of the deep, nearly gray Fe opacity. The spectrum therefore supports SiO but provides limited leverage on detailed silicate cloud structure and mineralogy; a physical condensate cloud remains consistent with the data but is less well constrained than for HD~89744B and GJ~1048B.

\subsection{Vertical SiO Depletion}
\label{subsec:sio_vertical}

Figure~\ref{fig:three_source_structure} compares the baseline Mg$_2$SiO$_4$+Fe+SiO retrievals for the three objects using the same physical cloud prescription. The Mg$_2$SiO$_4$ cloud base prior is defined relative to the equilibrium condensation pressure (Appendix~\ref{sec:clouds}), and the retrieved cloud bases remain close to the expected silicate condensation region. We use this chemically motivated reference to compare the vertical SiO profiles with the atmospheric pressures probed by the 4 and 8~$\mu$m bands. In equilibrium chemistry, SiO is an important Si-bearing gas in hot, deep layers, while Mg-bearing silicate condensation progressively removes silicon from the gas phase as the atmosphere cools \citep{2010ApJ...716.1060V}.

For each object, we compute the emission contribution functions with \texttt{petitRADTRANS} using the full opacity set from the Mg$_2$SiO$_4$+Fe+SiO retrieval. The contributions are averaged over 4.0--4.2 and 7.7--8.2~$\mu$m to trace the atmospheric pressures probed by the two spectral regions. The 4~$\mu$m overtone arises primarily from deeper layers, whereas the 7.7--8.2~$\mu$m core of the fundamental band probes lower pressures (Figure~\ref{fig:three_source_structure}). These lower pressure layers overlap more strongly with the region where equilibrium SiO is depleted by silicate condensation. As L dwarf atmospheres cool, the 8~$\mu$m band is therefore expected to become sensitive to SiO depletion before the deeper 4~$\mu$m overtone. This vertical structure is consistent with the observed mid-infrared sequence, in which SiO absorption persists through the early L types and weakens toward later L dwarfs as silicate condensation becomes increasingly important \citep{2026arXiv260715521B}.

The same vertical picture may also explain the gravity dependence reported by \citet{2026arXiv260715521B}, who found weaker SiO absorption in young, low gravity early L dwarfs than in field objects. To test this interpretation, we repeated the contribution function calculation for the physical cloud solutions after setting $\log g=4.5$, while keeping all other atmospheric and cloud parameters fixed. The lower gravity shifts the 7.7--8.2~$\mu$m contribution region to lower pressures, so the SiO band samples higher atmospheric layers where equilibrium SiO is more strongly depleted. This fixed parameter test isolates the direct effect of gravity: lowering $\log g$ shifts the 8~$\mu$m contribution to lower pressures, where SiO is more strongly depleted, consistent with the weaker SiO absorption observed in low gravity early L dwarfs. 

A related change with atmospheric temperature is seen within our sample. HD~89744B retains appreciable SiO to lower pressures than the two L1.5 companions, consistent with its warmer $P$--$T$ profile. Its SiO rich region consequently extends farther into the layers sampled by the 8~$\mu$m band, consistent with the strong local SiO preference measured for this object. In the cooler GJ~1048B and LSPM~J0632+5053B atmospheres, equilibrium SiO depletion begins at higher pressures, bringing the 8~$\mu$m contribution region closer to the depletion region.

\subsection{Deep Fe Cloud Structure}
\label{subsec:fe_clouds}

\begin{figure*}[!tp]
\centering
\includegraphics[width=\textwidth]{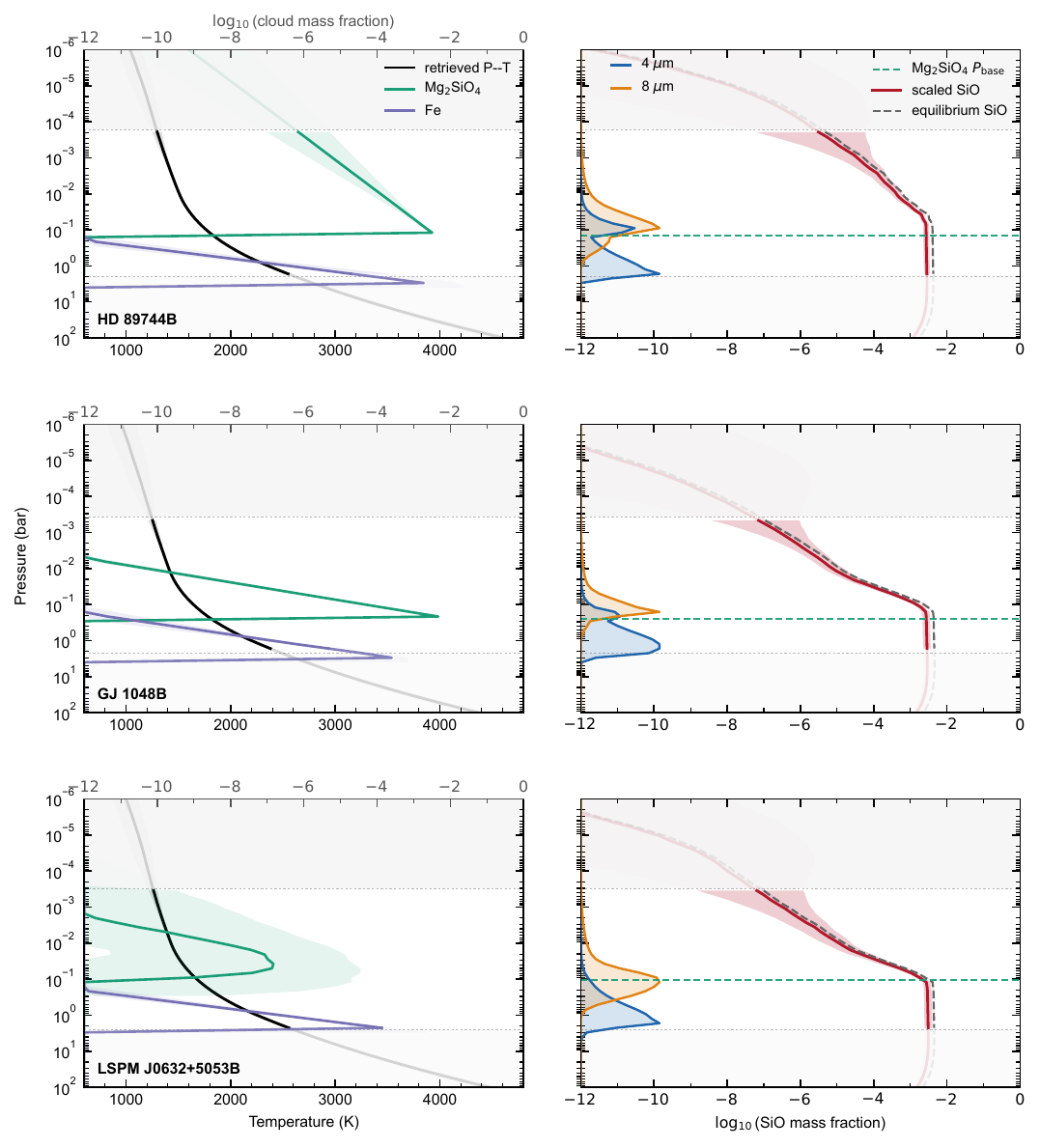}
\caption{
Vertical atmospheric structures from the baseline Mg$_2$SiO$_4$+Fe+SiO retrievals of HD~89744B (top), GJ~1048B (middle), and LSPM~J0632+5053B (bottom).
The left panels show the retrieved $P$--$T$ profiles and the Mg$_2$SiO$_4$ and Fe cloud mass fraction profiles; shaded bands denote the 16th--84th percentile ranges.
Gray regions indicate pressure ranges that contribute little to the observed emergent spectrum, where the retrieved thermal structure, cloud abundances, and gas abundance profiles are therefore weakly constrained.
The right panels compare the scaled and equilibrium SiO profiles and show the retrieved Mg$_2$SiO$_4$ cloud base pressures together with the normalized emission contribution functions for the 4.0--4.2 and 7.7--8.2~$\mu$m bands.
}
\label{fig:three_source_structure}
\end{figure*}

The physical cloud retrievals reveal a remarkably similar deep Fe cloud structure across the three objects (Figure~\ref{fig:three_source_structure}). The equilibrium Fe condensation curve intersects the retrieved $P$--$T$ profile at approximately 0.15~bar for HD~89744B and around 0.4~bar for GJ~1048B and LSPM~J0632+5053B. Despite these different condensation pressures, the retrieved Fe cloud bases converge to a narrow range of $\log P_{\rm base,Fe}\simeq0.46$--0.59, corresponding to approximately 3--4~bar. The Fe cloud base in HD~89744B therefore lies about 1.4~dex deeper than its equilibrium condensation point, while the offsets for the other two objects are about 0.9~dex. The retrieved sedimentation parameters are also consistently large, $f_{\rm sed,Fe}\simeq6.8$--8.2, indicating a rapid decrease in condensate abundance toward lower pressures and a vertically compact Fe cloud concentrated in the deep atmosphere \citep{2001ApJ...556..872A}.

Similar deep refractory opacity has been inferred across a broad range of cloudy substellar atmospheres. \citet{2017MNRAS.470.1177B} retrieved optically thick cloud decks near 5~bar in two mid L dwarfs and associated the opacity with corundum and/or Fe, while \citet{2021MNRAS.506.1944B} inferred an Fe deck near 10~bar in the L4.5 dwarf 2M2224$-$0158. Deep Fe clouds have also been retrieved for early T planetary mass analogs \citep{2023ApJ...944..138V}, the young planetary mass companion VHS~1256~b \citep{Whiteford2026}, and the early L companion $\kappa$~And~b \citep{2024ApJ...970...71X}. Our three objects extend this pattern by showing nearly identical Fe cloud base pressures even though their equilibrium condensation intersections occur at different atmospheric pressures. This recurring deep Fe structure suggests that the location of the retrieved Fe cloud base may remain near pressures of several bars across different substellar spectral types, with only a weak dependence on the pressure at which the atmospheric $P$--$T$ profile first crosses the equilibrium Fe condensation curve.

A finite evaporation region offers a physical explanation for Fe opacity extending below the equilibrium condensation boundary. Kinetic cloud calculations predict that settling particles can persist into hotter, undersaturated layers while gradually evaporating \citep{2004A&A...414..335W}. \citet{2021MNRAS.506.1944B} invoked the same process for 2M2224$-$0158, where molten Fe particles may continue to settle to higher pressures before fully vaporizing. The retrieved Fe component may also represent additional deep, nearly gray opacity that is not explicitly included in the adopted condensate set. The present spectra therefore support a compact deep opacity source, while its detailed physical origin remains uncertain.

\subsection{Atmospheric Composition and SiO Abundance Scaling}
\label{subsec:composition}

The quantities $[\mathrm{M/H}]{\rm gas}$ and $({\rm C/O}){\rm gas}$ in Table~\ref{tab:retrieval_results} are derived from the elemental abundances contained in the retrieved gas-phase species, including the scaled SiO abundance. Cloud condensates are excluded from this elemental inventory, so these quantities represent the composition of the modeled gas phase rather than the bulk elemental composition.

The preferred retrievals give
$({\rm C/O})_{\rm gas}=0.76^{+0.02}_{-0.03}$,
$0.89^{+0.02}_{-0.03}$, and
$0.82\pm0.02$
for HD~89744B, GJ~1048B, and LSPM~J0632+5053B, respectively.
The corresponding host-star C/O ratios are
$0.56\pm0.09$,
$0.69\pm0.12$, and
$0.57\pm0.09$
\citep{2026arXiv260709851P}.
Interestingly, the stellar and retrieved compositions show the same relative pattern: GJ~1048B has the highest C/O, while HD~89744B and LSPM~J0632+5053B have similar values.

Silicate condensation is expected to sequester oxygen into refractory condensates and increase the gas-phase C/O relative to the bulk composition. For the three host-star compositions, \citet{2026arXiv260709851P} estimate oxygen sink fractions of
$16.7\pm8.4\%$,
$21.8\pm9.5\%$, and
$25.7\pm4.3\%$.
Applying
$({\rm C/O})_{\rm gas,pred}=({\rm C/O})_{\star}/(1-f_{\rm O,sink})$
gives predicted gas-phase ratios of
$0.67\pm0.13$,
$0.88\pm0.19$, and
$0.77\pm0.13$,
respectively.
Notably, these values are consistent with the retrieved $({\rm C/O})_{\rm gas}$ for all three companions. The elevated gas-phase C/O ratios can therefore be explained by the host-star compositions with the expected sequestration of oxygen into refractory condensates.

In contrast, the preferred retrievals yield tightly clustered gas-phase metallicities of $[\mathrm{M/H}]_{\rm gas}=0.45$--$0.46$, systematically above both the host-star spectroscopic abundances ($[\mathrm{Fe/H}]\simeq0.19$--$0.25$) and photometric estimates ($[\mathrm{M/H}]_{\rm phot}=-0.02$--$0.23$), with the largest offset for LSPM~J0632+5053B \citep{2026arXiv260709851P}. The small posterior uncertainties in Table~\ref{tab:retrieval_results} reflect statistical precision within the adopted model, while the absolute metallicities remain sensitive to the retrieval framework, cloud treatment, thermal structure, observational uncertainties, and data set \citep{2024ApJ...976...82T,2025ApJS..281...49T,2025AJ....170...64Z,2025A&A...703A..79M,Whiteford2026}.

The preferred baseline models retrieve modestly sub-equilibrium SiO abundances for all three objects, with $\log f_{\rm SiO}$ between approximately $-0.1$ and $-0.2$. However, this common offset is sensitive to the assumed bulk metallicity and disappears in fixed-metallicity tests. These tests indicate that $\log f_{\rm SiO}$ is coupled to $[\mathrm{M/H}]$, C/O, and cloud opacity, and its absolute value also depends on the adopted equilibrium chemistry. We therefore interpret the retrieved $\log f_{\rm SiO}$ as a model-dependent correction to the equilibrium SiO profile rather than an absolute measure of SiO depletion.

\subsection{Silicate Cloud Mineralogy}
\label{subsec:mineralogy}

The broad mid-infrared silicate feature is sensitive to condensate composition, motivating additional physical-cloud retrievals with alternative Mg- and Si-bearing species while retaining SiO. For both HD~89744B and GJ~1048B, we test three additional silicate prescriptions relative to the baseline Mg$_2$SiO$_4$ model: MgSiO$_3$, MgSiO$_3$+SiO$_2$, and Mg$_2$SiO$_4$+MgSiO$_3$. The benchmark primaries provide an external reference: \citet{2026arXiv260709851P} measure Mg/Si ratios of $0.77\pm0.37$ for HD~89744 and $1.09\pm0.19$ for HD~16270, and associate their companions with enstatite+quartz and enstatite+forsterite clouds, respectively, based on the central values and phase-equilibrium predictions. All models considered here retain the same Fe cloud component, and $\Delta{\rm BIC}$ is defined relative to the best mineralogical model for each source, separately from the baseline comparisons in Table~\ref{tab:retrieval_results}.

For HD~89744B, Mg$_2$SiO$_4$ remains preferred among the four tested silicate prescriptions. MgSiO$_3$, Mg$_2$SiO$_4$+MgSiO$_3$, and MgSiO$_3$+SiO$_2$ give $\Delta{\rm BIC}=7.9$, 18.7, and 28.2, respectively. Thus, none of the tested two-silicate models improves the fit sufficiently to be favored over Mg$_2$SiO$_4$ alone. This differs from the enstatite+quartz composition inferred from the central host-star Mg/Si estimate, although the relatively large uncertainty in Mg/Si ($0.77\pm0.37$) weakens this comparison. More generally, the difference may also indicate that the observable cloud mineralogy depends on atmospheric thermal structure and condensation processes in addition to the bulk refractory composition.

For GJ~1048B, Mg$_2$SiO$_4$+MgSiO$_3$ gives the lowest BIC, while MgSiO$_3$+SiO$_2$, Mg$_2$SiO$_4$, and MgSiO$_3$ yield $\Delta{\rm BIC}=9.5$, 10.0, and 22.9. Models containing Mg$_2$SiO$_4$ are favored in both the single- and two-silicate comparisons. The preferred mixed model retrieves a larger cloud-base mass fraction for Mg$_2$SiO$_4$ than MgSiO$_3$, although differences in particle size and vertical distribution prevent these fractions from being interpreted directly as relative opacity contributions. The favored forsterite+enstatite composition is consistent with the Mg/Si-based expectation for HD~16270 \citep{2026arXiv260709851P}.

The contrast between the two objects is qualitatively consistent with the Mg-silicate condensation sequence: forsterite condenses at higher temperatures, while enstatite becomes more important as the atmosphere cools \citep{2010ApJ...716.1060V}. Accordingly, the warmer HD~89744B favors Mg$_2$SiO$_4$, whereas the cooler GJ~1048B favors Mg$_2$SiO$_4$+MgSiO$_3$. Although the sample and tested mineralogical set are too limited to establish a systematic sequence, the comparison suggests that both bulk composition and thermal structure contribute to the observable silicate mineralogy. Finally, we do not extend this analysis to LSPM~J0632+5053B because its spectrum ends at 9.3~$\mu$m and does not sufficiently constrain the full silicate feature (Section~\ref{subsec:lspm_results}).

\section{Conclusions}
\label{sec:conclusions}

We find evidence for gaseous SiO in all three benchmark early-L dwarfs from their broad JWST spectra. The clearest SiO signatures occur at 4.0--4.2 and 7.7--8.2~$\mu$m. Across 7.7--9.3~$\mu$m, the silicate cloud remains the dominant opacity source, while SiO contributes appreciable additional opacity throughout the same wavelength range. At $R=100$, variations in the silicate cloud opacity can reproduce much of the spectral structure beyond approximately 8.2~$\mu$m, increasing the degeneracy between the two opacity sources. The 7.7--8.2~$\mu$m region therefore retains the most distinctive SiO signature within the fundamental band. Across the viable cloud prescriptions, the 8~$\mu$m signature is recovered more consistently than the 4~$\mu$m overtone, which is more sensitive to the surrounding continuum shape.

The retrieved vertical structures connect these SiO signatures with the onset of silicate condensation. The 8~$\mu$m band probes lower pressures than the 4~$\mu$m band and therefore reaches atmospheric layers where equilibrium SiO is more strongly depleted by silicate condensation. As L dwarf atmospheres cool, SiO depletion therefore reaches the layers probed by the 8~$\mu$m band before the deeper 4~$\mu$m region. A fixed parameter test with lower $\log g$ further shifts the 8~$\mu$m contribution to lower pressures, where SiO is more strongly depleted, providing a natural explanation for the smaller SiO indices observed in low gravity early L dwarfs. The absolute SiO abundance scaling remains model dependent because of its covariance with metallicity, C/O, and cloud opacity. The physical cloud retrievals for all three sources also place compact Fe opacity near 3--4~bar, despite differences in the equilibrium Fe condensation pressures. The retrieved gas-phase C/O ratios agree with those expected from the host star compositions after accounting for refractory oxygen sequestration.

The spectra also provide constraints on silicate mineralogy. Using Mie scattering with amorphous silicate optical constants, HD~89744B favors Mg$_2$SiO$_4$, whereas GJ~1048B favors a mixed Mg$_2$SiO$_4$+MgSiO$_3$ cloud. The evidence of gaseous SiO in these early-L atmospheres provides a direct observational tracer of the silicon reservoir immediately preceding silicate condensation. Extending the same analysis to cooler L dwarfs with complete coverage of the 4~$\mu$m overtone and the 8--11~$\mu$m silicate feature will allow the disappearance of gaseous SiO, the evolution of silicate mineralogy, and the predictions of thermochemical condensation models to be tested along the L-dwarf sequence.

\section*{Data Availability}

The JWST observations analyzed in this work are publicly available from the Mikulski Archive for Space Telescopes (MAST) under GO program 3670 (PI B. Burningham), and can be accessed via \dataset[doi:10.17909/39ps-kr46]{\doi{10.17909/39ps-kr46}}. The atmospheric retrieval diagnostic products presented in this work are available on Zenodo:
\dataset[doi:10.5281/zenodo.22200294]{https://doi.org/10.5281/zenodo.22200294}.

\begin{acknowledgments}
This work is supported by the National Natural Science Foundation of China (NSFC) through projects 12588202, 12622304, 12603059, 12373028, 12633007, and 12322306. S.W. and X.C. acknowledge support from the Youth Innovation Promotion Association of the Chinese Academy of Sciences under grant Nos. 2023065 and 2022055, respectively. Z.T. acknowledges support from the Postdoctoral Fellowship Program of CPSF under Grant Number GZC20261725. This work makes use of observations obtained with the NASA/ESA/CSA James Webb Space Telescope. The JWST data were retrieved from the Mikulski Archive for Space Telescopes (MAST) at the Space Telescope Science Institute (STScI). STScI is operated by the Association of Universities for Research in Astronomy, Inc. (AURA), under NASA contract NAS5-03127 for JWST. The observations used in this work are associated with JWST programs 3670 (PI B. Burningham).
\end{acknowledgments}





\facilities{JWST (NIRSpec/PRISM, MIRI/LRS)}

\software{JWST Calibration Pipeline (v3.0.0), CRDS (v13.2.7; \citealt{2016A&C....16...41G}), \texttt{SEDkit} \citep{2015ApJ...810..158F}, \texttt{petitRADTRANS} \citep{2019A&A...627A..67M,2024JOSS....9.5875N}, \texttt{Exo\_k} \citep{2021A&A...645A..20L}, PyMultiNest/MultiNest \citep{2009MNRAS.398.1601F}}

\restartappendixnumbering
\appendix

\section{Detailed Atmospheric Retrieval Framework}
\label{app:retrieval_framework}

\subsection{Radiative Transfer and Opacity Sources}
\label{sec:radiative_transfer}

We computed one-dimensional emission spectra in correlated-$k$ mode. The opacity tables were rebinned with \texttt{Exo\_k} for the retrieval calculations \citep{2021A&A...645A..20L}.

The molecular and atomic line opacity sources included Na \citep{2019A&A...628A.120A}, K \citep{2019A&A...627A..67M}, H$_2$O \citep{2018MNRAS.480.2597P}, CH$_4$ \citep{2020ApJS..247...55H}, CO \citep{2010JQSRT.111.2139R}, CO$_2$ \citep{2020MNRAS.496.5282Y}, NH$_3$ \citep{2019MNRAS.490.4638C}, PH$_3$ \citep{2015MNRAS.446.2337S}, H$_2$S \citep{2016MNRAS.460.4063A}, FeH \citep{2010A&A...523A..58W}, VO \citep{2016MNRAS.463..771M}, and TiO \citep{2019MNRAS.488.2836M}. SiO opacity \citep{2022MNRAS.510..903Y} was added only to the with-SiO retrievals.

Continuum opacity included Rayleigh scattering by H$_2$ \citep{1962ApJ...136..690D} and He \citep{1965PPS....85..227C}, together with collision-induced absorption from H$_2$--H$_2$ \citep{2001JQSRT..68..235B,2002A&A...390..779B} and H$_2$--He \citep{1988ApJ...326..509B,1989ApJ...336..495B,1989ApJ...341..549B}.

\subsection{Pressure--Temperature Structure}\label{sec:pt_profile}

Following \citet{2025AJ....170...64Z}, we parameterize the pressure--temperature ($P$--$T$) profile using the temperature at the bottom boundary and nodes in the logarithmic thermal gradient, $\nabla \equiv d\ln T/d\ln P.$ We adopt $T_{\rm bottom}\equiv T(100~{\rm bar})$ to set the absolute temperature scale and sample five thermal gradients, $\nabla_1$--$\nabla_5$, at pressures of 100, 1, $10^{-2}$, $10^{-4}$, and $10^{-6}$~bar, respectively. The atmospheric structure is evaluated on 64 pressure layers spaced uniformly in $\log P$ from $10^{-6}$ to $10^2$~bar. We interpolate the thermal gradient linearly in $\log P$ between the five nodes and integrate the profile upward from $T_{\rm bottom}$.

We use cloudy $P$--$T$ profiles from the Sonora Diamondback grid to guide the priors on the thermal gradients \citep{2024ApJ...975...59M}. The ensemble contains 150 equally weighted profiles spanning $T_{\rm eff}=2000$--$2400$~K in 100~K steps, $\log g=5.0$ and 5.5, $[\mathrm{M/H}]=-0.5$, 0.0, and $+0.5$, and $f_{\rm sed}=1$, 2, 3, 4, and 8. For each profile, we calculate $\nabla$ on the native pressure grid and interpolate it in $\log P$ to the retrieval-node pressures. The resulting gradient distributions guide the choice of truncated-normal priors for $\nabla_1$--$\nabla_3$. The $10^{-4}$ and $10^{-6}$ bar nodes extend beyond the pressure range directly sampled by the Sonora profiles, so we adopt independent nonnegative uniform priors for $\nabla_4$ and $\nabla_5$. The numerical prior ranges are listed in Table~\ref{tab:retrieval_parameters}.

\begin{deluxetable}{ll}
\tablecaption{Retrieval Parameters and Priors\label{tab:retrieval_parameters}}
\tabletypesize{\scriptsize}
\tablewidth{0pt}
\tablehead{
\colhead{Sampled Parameter} &
\colhead{Prior}
}
\startdata
\cutinhead{Shared baseline}
$T_{\rm bottom}$ (K) & $\mathcal{U}(1200,6000)$ \\
$\nabla_1$ & $\mathcal{TN}(0.14,0.02;0.10,0.20)$ \\
$\nabla_2$ & $\mathcal{TN}(0.145,0.045;0.01,0.22)$ \\
$\nabla_3$ & $\mathcal{TN}(0.075,0.04;0.035,0.155)$ \\
$\nabla_4$ & $\mathcal{U}(0,0.10)$ \\
$\nabla_5$ & $\mathcal{U}(0,0.10)$ \\
$[\mathrm{M/H}]$ & $\mathcal{U}(-0.5,0.7)$ \\
C/O & $\mathcal{U}(0.1,1.5)$ \\
$\log g$ & $\mathcal{N}(5.33,0.10)$ \\
$R/R_{\rm Jup}$ & $\mathcal{N}(0.97,0.08)$ \\
$10^{b_{\rm NIR}}$ & $\mathcal{U}\!\left[(0.1\sigma_{{\rm NIR},\min})^2,(10\sigma_{{\rm NIR},\max})^2\right]$ \\
$10^{b_{\rm MIRI}}$ & $\mathcal{U}\!\left[(0.1\sigma_{{\rm MIRI},\min})^2,(10\sigma_{{\rm MIRI},\max})^2\right]$ \\
\cutinhead{SiO increment}
$\log f_{\rm SiO}$ & $\mathcal{U}(-3,3)$ \\
\cutinhead{gray cloud increment}
$\log\kappa_{\rm gray}$ (cm$^2$ g$^{-1}$) & $\mathcal{U}(-6,2)$ \\
$\log P_{\rm cloud}$ (bar) & $\mathcal{U}(-6,2)$ \\
$\gamma_{\rm cloud}$ & $\mathcal{U}(0,10)$ \\
\cutinhead{Physical cloud increment}
$\log X_{0,s}$ & $\mathcal{U}(-12,0)$ \\
$\Delta\log P_{{\rm base},s}$$^{(a)}$ & $\mathcal{U}(-1,1)$ \\
$f_{{\rm sed},s}$ & $\mathcal{U}(0,10)$ \\
$\log r_s$ ($\mu$m) & $\mathcal{U}(-2,2)$ \\
$\sigma_g$$^{(b)}$ & $\mathcal{U}(1.02,3)$ \\
\enddata
\tablecomments{
$\mathcal{U}(a,b)$ denotes a uniform prior on $[a,b]$, 
$\mathcal{N}(\mu,\sigma)$ denotes a normal distribution with mean $\mu$ and standard deviation $\sigma$, 
and $\mathcal{TN}(\mu,\sigma;a,b)$ denotes the corresponding normal distribution truncated to $[a,b]$. (a) For the Fe cloud of HD~89744B, we adopt a broader prior of $\mathcal{U}(-2,2)$ to allow the cloud base to extend beyond the range permitted by the default prior. (b) The lognormal particle-size width $\sigma_g$ is shared among all condensate species within a given physical cloud retrieval.
}
\end{deluxetable}

\subsection{Equilibrium Chemistry and SiO Hypotheses}\label{sec:chemistry}

Except for SiO, the gas composition is assumed to be in chemical equilibrium and is parameterized by metallicity, $[\mathrm{M/H}]$, and the elemental carbon-to-oxygen ratio, C/O. At each atmospheric layer, we interpolate the equilibrium mass fractions from the expanded chemistry grid\footnote{\url{https://doi.org/10.5281/zenodo.15654694}} described by \citet{2025AJ....170...64Z}. The grid spans temperatures of 40--6000~K, pressures of $10^{-8}$--$10^{3}$~bar, metallicities of $-2$ to $+3$~dex, and C/O ratios of 0.1--1.6.

Measurements of the coeval primaries indicate near-solar compositions (Table~\ref{tab:observations}; \citealt{2026arXiv260709851P}). We therefore adopt a common metallicity prior of $[\mathrm{M/H}]\sim\mathcal{U}(-0.5,0.7)$ for all three companions. Here, $[\mathrm{M/H}]$ and C/O denote the elemental input parameters used to interpolate the equilibrium chemistry grid.

We consider two hypotheses for gaseous SiO. In the with-SiO retrievals, the equilibrium SiO profile is scaled by a single free abundance factor,
\begin{equation}
X_{\rm SiO}(P)
=
10^{\log f_{\rm SiO}}
X_{\rm SiO,eq}(P),
\label{eq:sio_scaling}
\end{equation}
where $X_{\rm SiO,eq}(P)$ is the equilibrium SiO mass fraction along the atmospheric $P$--$T$ profile and $\log f_{\rm SiO}$ is a global logarithmic scaling factor. We adopt
$\log f_{\rm SiO}\sim\mathcal{U}(-3,3)$.
In the no-SiO retrievals, we set $X_{\rm SiO}=0$ and exclude both the SiO line opacity and the $\log f_{\rm SiO}$ parameter.

\subsection{Cloud Prescriptions}
\label{sec:clouds}

We use three baseline cloud prescriptions: cloud-free, gray, and physical condensate clouds. The cloud-free and gray models provide limiting treatments of condensate opacity, while the physical cloud models include wavelength-dependent absorption and scattering by individual condensate species. The baseline physical cloud composition consists of Mg$_2$SiO$_4$ and Fe.

The gray cloud prescription introduces a wavelength-independent opacity with a pressure-dependent vertical profile,
\begin{equation}
\kappa_{\rm gray}(P)
=
\kappa_0
\left[
\min\left(\frac{P}{P_{\rm cloud}},1\right)
\right]^{\gamma_{\rm cloud}},
\label{eq:gray_cloud}
\end{equation}
where $\kappa_0$ is the deep cloud opacity reached at $P\geq P_{\rm cloud}$, $P_{\rm cloud}$ sets the pressure above which the opacity decreases, and $\gamma_{\rm cloud}$ controls the vertical decline.

For each condensate species $s$ in the physical cloud models, we parameterize the cloud by its mass fraction at the cloud base, $X_{0,s}$, and the cloud base pressure, $P_{{\rm base},s}$. Above the cloud base, the condensate mass fraction follows
$X_s(P)=10^{\log X_{0,s}}(P/P_{{\rm base},s})^{f_{{\rm sed},s}}$
for $P\leq P_{{\rm base},s}$ and is set to zero at higher pressures \citep{2001ApJ...556..872A}. The sedimentation parameter $f_{{\rm sed},s}$ controls how rapidly the condensate abundance decreases toward lower pressures. Each species has independent values of $\log X_{0,s}$, $f_{{\rm sed},s}$, and the characteristic particle radius $\log r_s$. We assume a lognormal particle size distribution with a width $\sigma_g$ shared by all cloud species.

Cloud opacity is calculated with Mie theory using amorphous optical constants for the silicate species. Previous modeling and retrieval studies have reproduced mid-infrared silicate features with small amorphous grains of forsterite, enstatite, and quartz \citep{2021MNRAS.506.1944B,2021ApJ...920..146L,2023MNRAS.523.4739S,Whiteford2026}. We use the same optical treatment for all physical cloud models so that differences between the mineralogical retrievals primarily reflect the adopted condensate composition.

For the baseline Mg$_2$SiO$_4$+Fe model, we adopt the equilibrium condensation curve of \citet{2010ApJ...716.1060V} for Mg$_2$SiO$_4$ and the Fe saturation relation of \citet{2001ApJ...556..872A}. For each condensate, $P_{{\rm cond},s}$ is defined by the intersection of the sampled $P$--$T$ profile with the corresponding condensation curve or saturation relation. We then use $P_{{\rm cond},s}$ to set the prior on the cloud base pressure rather than sampling the cloud base independently of the condensation chemistry. Specifically, we define
$\Delta\log P_{{\rm base},s}
\equiv
\log P_{{\rm base},s}-\log P_{{\rm cond},s}$,
with a default prior of
$\Delta\log P_{{\rm base},s}\sim\mathcal{U}(-1,1)$.
This allows the cloud base to vary by $\pm1$ dex around the equilibrium condensation pressure.

We also explore alternative silicate compositions containing MgSiO$_3$, Mg$_2$SiO$_4$+MgSiO$_3$, and MgSiO$_3$+SiO$_2$, using the corresponding condensation relations from \citet{2001ApJ...556..872A} and \citet{2023ApJ...956L..32G}. Apart from the condensate composition and its condensation relation, these retrievals use the same cloud parameterization and amorphous Mie treatment as the baseline physical cloud model.

\subsection{Global Parameters and Likelihood}
\label{sec:likelihood}

The surface gravity and radius are free parameters in all retrievals. Based on the evolutionary model predictions in Section~\ref{sec:evolutionary_parameters} across $[\mathrm{M/H}]=-0.5$, 0.0, and $+0.5$, we adopt Gaussian priors of $\log g\sim\mathcal{N}(5.33,0.10)$ and $R/R_{\rm Jup}\sim\mathcal{N}(0.97,0.08)$ for all retrievals.

The modeled surface flux is scaled to the observed flux by $(R/d)^2$, where $R$ is the retrieved radius and $d$ is fixed by the measured parallax. We require the corresponding mass, $M=gR^2/G$, to satisfy $M\leq120\,M_{\rm Jup}$, limiting combinations of surface gravity and radius that imply masses substantially above the expected range for these early-L companions.

We include independent additive variance terms for the NIRSpec and MIRI spectra. For each spectral point, the effective variance is $s_i^2=\sigma_i^2+10^{b_{k(i)}}$, where $\sigma_i$ is the measured flux density uncertainty and $k(i)$ denotes the NIRSpec region below 5.3~$\mu$m or the MIRI region at and above 5.3~$\mu$m. For each region, we adopt $10^{b_k}\sim\mathcal{U}[(0.1\sigma_{k,\min})^2,(10\sigma_{k,\max})^2]$, corresponding to a prior uniform in the added variance.

Assuming independent Gaussian residuals, the log likelihood is
\begin{equation}
\ln\mathcal{L}
=
-\frac{1}{2}
\sum_i
\left[
\frac{(F_i-M_i)^2}{s_i^2}
+
\ln\left(2\pi s_i^2\right)
\right],
\label{eq:likelihood}
\end{equation}
where $F_i$ and $M_i$ are the measured and modeled flux densities, respectively.

We sampled each retrieval with MultiNest through the \texttt{PyMultiNest} interface \citep{2009MNRAS.398.1601F}. The standard runs used 1000 live points, constant efficiency mode, a sampling efficiency of 0.05, and an evidence tolerance of 0.5. For selected higher dimensional cloud models, we reduced the sampling efficiency to 0.02 to improve the robustness of posterior exploration. The retrieval parameters and priors are summarized in Table~\ref{tab:retrieval_parameters}. 


\end{document}